\documentclass{easychair}

\usepackage{doc}
\usepackage{doi}

\usepackage{graphicx}
\usepackage{hyperref}

\usepackage[utf8x]{inputenc}
\usepackage{amsmath}
\usepackage{amssymb}
\usepackage{upgreek}
\usepackage{xcolor}
\usepackage{colortbl}
\usepackage{enumitem}

\usepackage{tikz}
\usetikzlibrary{arrows,positioning,calc,shapes.geometric}
\usepackage{tabularx}

\definecolor{keywordcolor}{rgb}{0.7, 0.1, 0.1}   
\definecolor{tacticcolor}{rgb}{0.1, 0.2, 0.6}    
\definecolor{commentcolor}{rgb}{0.4, 0.4, 0.4}   
\definecolor{symbolcolor}{rgb}{0.0, 0.1, 0.6}    
\definecolor{sortcolor}{rgb}{0.1, 0.5, 0.1}      

\usepackage{listings}
\usepackage{flushend}
\usepackage{xspace}
\usepackage{underscore}

\newcommand{\QQ}{\mathbb{Q}}
\newcommand{\Qp}{\mathbb{Q}_p}
\newcommand{\ZZ}{\mathbb{Z}}
\newcommand{\Zp}{\mathbb{Z}_p}
\newcommand{\NN}{\mathbb{N}}

\newcommand{\mathlib}{\textsf{mathlib}\xspace}
\newcommand{\move}{\textsf{move}\xspace}
\newcommand{\elim}{\textsf{elim}\xspace}
\newcommand{\squash}{\textsf{squash}\xspace}

\newcommand{\lean}[1]{\lstinline[language=lean]{#1}}
\newcommand{\HC}[1]{\ensuremath{\mathcal{H}(\text{\lean{#1}})}}
\newcommand{\IC}[1]{\ensuremath{\mathcal{I}(\text{\lean{#1}})}}

\usepackage{scalefnt}

\title{Simplifying Casts and Coercions%
\thanks{We acknowledge support
from the European Research Council (ERC) under
the European Union's Horizon 2020 research and innovation
program (grant agreement No. 713999, Matryoshka) 
and from the Dutch Research Council (NWO) under the
Vidi program (project No. 016.Vidi.189.037, Lean Forward).}}

\author{
    Robert Y.\ Lewis\inst{1}
\and
    Paul-Nicolas Madelaine\inst{2}
}

\institute{
  Vrije Universiteit Amsterdam,
  The Netherlands\\
  \email{r.y.lewis@vu.nl}
\and  
   \'Ecole Normale Sup\'erieure,
   Paris, France\\
   \email{paul-nicolas.madelaine@ens.fr}
}

\authorrunning{Lewis and Madelaine}

\titlerunning{Simplifying Casts and Coercions}

\begin{document}

\maketitle

\begin{abstract}
  This paper introduces \lean{norm_cast}, 
  a toolbox of tactics for the Lean proof assistant 
  designed to manipulate expressions containing coercions and casts.
  These expressions can be frustrating for beginning and expert users alike;
  the presence of coercions can cause seemingly identical expressions to fail to unify
  and rewrites to fail.
  The \lean{norm_cast} tactics aim to make reasoning with such expressions as transparent as possible.
  They are used extensively to eliminate boilerplate arguments in the Lean mathematical library
  and in external developments.
\end{abstract}




\section{Introduction}
\label{section:intro}

Many popular type-theoretic foundations for proof assistants,
including the Calculus of Inductive Constructions,
do not have native subtypes. 
Even for numeric types like $\NN$, $\ZZ$, and $\QQ$
with a natural chain of inclusions, 
terms must be cast from one to another with an explicit function application.
The numeral \lean{5 : ℕ} is syntactically different from \lean{5 : ℤ} and \lean{5 : ℝ}. 
To construct the sum of variables \lean{n : ℕ} and \lean{z : ℤ},
one needs either an unwieldy sum operator with type \lstinline{ℕ → ℤ → ℤ}
or a way to lift \lean{n} to the larger type \lstinline{ℤ}.

Inserting coercions is a common programming language feature,
and proof assistants are no exception: 
many modern systems will interpret \lean{n + z} in a reasonable way.
Combined with type-polymorphic operations and relations like $+$ and $<$
and generic numeral expressions, 
subtyping concerns can largely be ignored at the input level.
However, the ease of input often belies the complexity of the underlying term.
Using such terms in practice can go against the grain of intuition,
especially for users coming from mathematics,
where one almost never makes such distinctions.
It is frustrating to realize that work must be done
to unify \lstinline{n < (5 : ℕ)} with \lstinline{↑n < (5 : ℤ)},
where \lstinline{↑n} denotes the cast of \lean{n} into \lstinline{ℤ}.

A more intricate example of this frustration 
appears in the Lean development of the $p$-adic numbers~\cite{Lewi19}
while proving 
\begin{lstlisting}
theorem of_int {p : ℕ} (hp : prime p) (z : ℤ) : padic_norm p ↑z ≤ 1
\end{lstlisting}
where \lstinline{padic_norm : ℕ → ℚ → ℚ}. 
Straightforward manipulation reduces the proof to three goals: 
\begin{itemize}
  \item \lstinline{prime p ⊢ (1 : ℤ) ≤ ↑p}
  \item \lstinline{z ≠ (0 : ℤ) ⊢ -padic_val_rat p ↑z ≤ (0 : ℤ)}
  \item \lstinline{z ≠ (0 : ℤ) ⊢ ↑z ≠ (0 : ℚ)}
\end{itemize}
To solve these goals by hand, 
the user must combine knowledge of library lemmas with lemmas that manipulate casts. 
The latter obscure the main ideas of the proof:
\begin{lstlisting}
{ rw [←nat.cast_one, nat.cast_le], exact le_of_lt hp.one_lt },
{ rw [padic_val_rat_of_int _ hp.ne_one hz, neg_nonpos],
  exact int.coe_nat_nonneg _ },
{ exact int.cast_ne_zero.2 hz }
\end{lstlisting}

We introduce a family of tactics 
implemented in the Lean proof assistant~\cite{Mour15a} 
that aim to remove these frustrations.
The core tool, \lean{norm_cast}, 
tries to rewrite an expression containing casts
to a normal form determined by a configurable collection of rewrite rules.
Variants allow the user to apply lemmas and hypotheses 
and rewrite the goal 
``modulo'' the presence of casts.
The tool was developed to address usability issues 
raised while formalizing mathematical results in Lean\footnote{
  Lean users previously wrote a guide to managing casts by hand.
  This guide is archived with our supplementary materials.}~\cite{Dahm19,Lewi19}.
it is incorporated into Lean's mathematical library \mathlib~\cite{mathlib20}, 
where it is already invoked 372 times, 
and is also used heavily in external libraries~\cite{BCM20}. 

Using our tool, the script above focuses on the relevant library lemmas:
\begin{lstlisting}
{ exact_mod_cast le_of_lt hp.one_lt },
{ rw [padic_val_rat_of_int _ hp.ne_one hz, neg_nonpos],
  norm_cast; simp },
{ exact_mod_cast hz }
\end{lstlisting}

Our tool is extensible: 
adapting it to new theories with new coercions simply requires tagging certain library lemmas.
It is not restricted to concrete types like $\NN$, $\ZZ$, and $\QQ$:
it supports casts into abstract algebraic structures, casts out of arbitrary subtypes,
and in general all casts that are well behaved.
It is largely independent of the underlying logic:
for example, there are no roadblocks to implementing it in systems without convertibility.

We provide a website\footnote{\url{https://lean-forward.github.io/norm_cast}} 
which points to our code in the \mathlib repository, along with examples of \lean{norm_cast} in use.
Our aim in this report is \emph{not} to give a theoretical account of type inclusions,
but rather to present a powerful and lightweight procedure 
that is effective at dealing with these situations in practice.

``Coercion'' and ``cast'' are sometimes used interchangeably in the literature, 
and ``cast'' can also refer to the transport of a term \lean{t : A} 
to the type \lean{B} along a type equality \lean{A = B}.
In this description, we take a \emph{cast} \lstinline{↑ : A → B} to be simply a function; 
it typically preserves structure, and is often injective, but neither is required. 
\emph{Casting} a term \lean{t : A} to \lean{B} 
refers to applying the (often canonical) cast. 
A \emph{coercion} is a cast that is automatically inserted by the elaborator.
We do not consider casts along type equalities.

\section{Lean Specifics}
\label{section:lean}

While the approach we describe can be adapted to other proof assistants, 
some details of this report are specific to Lean. 
Here we describe some of the relevant features of Lean.

Lean's elaborator inserts coercions using type classes~\cite{Sels20,Wadl89}. 
Its generic coercion function has signature
\begin{lstlisting}
coe : Π {a : Sort u} {b : Sort v} [has_lift_t a b], a → b
\end{lstlisting}
where the arguments \lean{a} and \lean{b} are implicit 
and the anonymous \lean{has_lift_t} argument is inferred by type class resolution. 
An instance of the type class \lean{has_lift_t a b} 
witnesses a transitive chain of coercions from \lean{a} to \lean{b}, 
avoiding loops caused by reflexive instances. 
When a function application fails to typecheck, 
the elaborator will insert applications of \lean{coe} 
and try to resolve the resulting \lean{has_lift_t} goal.
Coercions are typically inserted at the leaf nodes of an expression.
Users can also manually insert casts by using \lean{coe} directly, 
with prefix notation \lstinline{↑}.

Type-polymorphic operators and relations like \lean{+} and \lean{<} 
are also implemented with type classes. 
Numerals build on top of these. 
A numeral is represented in binary by nested applications of the following terms:
\begin{lstlisting}
zero : Π (α : Type u) [has_zero α], α
one  : Π (α : Type u) [has_one α], α
bit0 : Π {α : Type u} [has_add α], α → α
bit1 : Π {α : Type u} [has_one α] [has_add α], α → α
\end{lstlisting}
Any type \lstinline{α} that instantiates the classes 
\lean{has_zero α}, \lean{has_one α}, and \lean{has_add α} 
supports numeral notation, e.g.\ \lean{(5 : α)}. 
While in this description we explicitly write the types of all numerals, 
in practice they are typically inferred.

Lean's powerful metaprogramming framework~\cite{EURAM17} allows us 
to implement our tool in the language of Lean itself 
and include it in \mathlib. 
The framework provides an interface to access the system's routines 
for unification, type class resolution, simplification, and more. 
Metaprograms can query and add to a Lean environment. 
Declarations in an environment can be tagged with parametrized \emph{attributes}, 
and metaprograms are able to define new attributes, 
use them to tag declarations, 
and retrieve lists of tagged declarations.

\section{Outline of the Simplification Procedure}
\label{section:outline}

The core routine in our procedure 
takes as input an expression 
and transforms the expression to one in which 
applications of the cast function \lstinline{↑} are normalized. 
It returns a proof that the resulting expression is equal to the input.
In the most common case, where the expression is a proposition, 
the proof of equality serves as a proof of logical equivalence.

As an example of the expected behavior, 
we simplify the expression \lstinline{↑m + ↑n < (10 : ℤ)}, 
where \lean{m, n : ℕ} are cast to \lstinline{ℤ}. 
The expected normal form is \lean{m + n < (10 : ℕ)}; 
recall that \lstinline{+}, \lstinline{<}, and \lstinline{10} are polymorphic. 
Our tool should proceed as follows:
\begin{enumerate}
\itemsep1\jot
  \item \label{step:outline:numeral} 
    Replace the numeral on the right with the cast of a \lean{nat}: \lstinline{↑m + ↑n < ↑(10 : ℕ)}
  \item \label{step:outline:move} 
    Factor \lstinline{↑} to the outside of the left: \lstinline{↑(m + n) < ↑(10 : ℕ)}
  \item \label{step:outline:elim} 
    Eliminate both casts to get an inequality over \lstinline{ℕ}: \lstinline{m + n < (10 : ℕ)}
\end{enumerate}

Steps~\ref{step:outline:move} and \ref{step:outline:elim} 
are effectively just applications of Lean's simplification API with certain rewrite lemmas. 
Step \ref{step:outline:numeral} has a slightly different flavor, 
but we will still be able to use the simplification API to implement this. 
Since the simplifier will handle cases of these patterns nested inside larger expressions, 
we can focus on the atomic situation.

Each of these steps must be justified by lemmas in the library, of course. 
They would not be sound for arbitrary types, operations, and relations. 
Users of our tool tag certain declarations with the attribute \lean{norm_cast}, for example:
\begin{lstlisting}
@[norm_cast] 
theorem nat.cast_add {α : Type} [add_monoid α] [has_one α] (m n : ℕ) : 
  (↑(m + n) : α) = ↑m + ↑n := ...
\end{lstlisting}
Our tool sorts these tagged declarations into three categories.

\begin{itemize}
\itemsep1\jot

\item \move lemmas equate expressions with casts at the root 
  to expressions with casts further toward the leaves, 
  e.g.\ \lstinline{↑(m + n) = ↑m + ↑n}.
  By \mathlib convention, 
  such lemmas are stated with the root cast on the left of the equation; 
  Step~\ref{step:outline:move} uses them as right-to-left rewrite rules.

\item \elim lemmas relate expressions with casts 
  to expressions without casts, 
  e.g.\ \lstinline{↑a < ↑b} \lstinline{↔} \lstinline{a < b}. 
  Such lemmas are stated with the expression containing casts on the left of the relation; 
  Step~\ref{step:outline:elim} uses them as left-to-right rewrite rules. 
  These lemmas are not restricted to propositional equivalences: 
  they can also be used to modify polymorphic operations, 
  e.g.\ \lstinline{∥↑a∥ = ∥a∥} for a real valued norm function defined on all normed spaces.

\item \squash lemmas equate expressions with one or more casts at the root 
  to expressions with fewer casts at the root, 
  e.g.\ \lstinline{↑(1 : ℕ) = (1 : ℤ)} and \lstinline{↑↑n = ↑n}. 
  Such lemmas are stated with the expression containing the larger number of casts on the left; 
  Step~\ref{step:outline:numeral} uses them alongside \move lemmas 
  to justify that \lstinline{(10 : ℤ) = ↑(10 : ℕ)}, 
  and they are used in the heuristic splitting step described below.

\end{itemize}

To simplify expressions where casts come from a variety of sources, 
we must sometimes split casts into pieces. 
Suppose \lstinline{n : ℕ} and \lstinline{z : ℤ}, 
and consider the goal \lstinline{↑n + ↑z = (2 : ℚ)}. 
(We call the pattern \lstinline{P (↑x) (↑y)}, 
where \lean{P} is a binary function or relation taking two arguments of the same type 
and \lean{x} and \lean{y} are of different types, 
the \emph{heuristic splitting pattern}.) 
We cannot rewrite the left hand side to \lstinline{↑(n + z)}, 
since the addition would not be well typed.
However, \move and \squash lemmas justify a transformation to \lstinline{↑↑n + ↑z = ↑↑(2 : ℕ)}, 
where the inner casts go \lstinline{ℕ → ℤ} and the outer \lstinline{ℤ → ℚ}. 
We transform this to \lstinline{↑(↑n + z) = ↑↑(2 : ℕ)} and then \lstinline{↑n + z = ↑(2 : ℕ)}. 
Finally, \squash lemmas reduce the right hand side to the native numeral \lstinline{(2 : ℤ)}.

\section{Implementation}
\label{section:implementation}

The core simplification routine has type \lean{expr → tactic (expr × expr)}, 
taking in an expression and returning it in normal form 
with a proof that the output is equal to the input. 
Lean's simplifier API provides methods for traversing and rewriting an expression 
from the leaf nodes outward (``bottom up'') and in reverse (``top down''). 
Our routine consists of four successive simplifier passes.

\begin{enumerate}
\itemsep1\jot
  \item \label{step:impl:numeral} 
    Working top down, replace each numeral \lstinline{(num : α)} with \lstinline{↑(num : ℕ)}. 
    Justify these replacements with \move lemmas 
    to move casts down through applications of the constants \lean{bit0} and \lean{bit1}, 
    and \squash lemmas 
    to change \lstinline{↑(0 : ℕ)} and \lstinline{↑(1 : ℕ)} to \lstinline{(0 : α)} and \lstinline{(1 : α)}.
 
  \item \label{step:impl:mvel} 
    Working bottom up, move casts upward by rewriting with \move lemmas 
    and eliminate them when possible by rewriting with \elim lemmas. 
    If no rewrite rules apply to a subexpression that matches the heuristic splitting pattern, 
    fire the \emph{splitting procedure} described below. 
 
  \item \label{step:impl:squash} 
    Working top down, clean up any unused repeated casts that were inserted by the heuristic 
    by rewriting with \squash lemmas.
 
  \item \label{step:impl:revnum} 
    Working top down, restore numerals to their natively typed form as in Step~\ref{step:impl:numeral}. 
    This is again justified by \move and \squash lemmas.
\end{enumerate}

The splitting procedure fires on an expression of the form \lstinline{P (↑x) (↑y)}, 
where \lean{P} is a binary function or relation, 
\lean{x : X} and \lean{y : Y} are both cast to type \lean{Z}, 
and \lean{X} and \lean{Y} are not equal. 
The procedure tries to find a coercion from \lean{X} to \lean{Y} or vice versa. 
The existence of a coercion is expressed as a type class instance, 
so this can be tested by trying to resolve a type class goal \lean{has_lift_t X Y}. 
Supposing the former coercion is found, 
the procedure tries to replace \lstinline{↑x} with \lstinline{↑↑x}, 
where the nested coercions go from \lean{X} to \lean{Y} to \lean{Z}. 
This is justified using \squash lemmas. 

We use Lean's user attribute API to define an attribute \lean{norm_cast}. 
This attribute is applied by the user to a lemma at or after the time of declaration. 
It tags the lemma for use in the procedure. 
A \lean{norm_cast} lemma has the form \lean{lhs = rhs} or \lean{lhs ↔ rhs}, 
typically preceded by a sequence of quantifiers. 
In nearly all cases, 
the attribute handler can automatically classify a lemma as \elim, \move, or \squash. 

\emph{Head casts} are applications of casts that appear at the root of the expression tree, 
as in \lstinline{↑↑(x+y)}, 
and \emph{internal casts} appear elsewhere. 
Let \HC{e} and \IC{e} denote the number of head casts and internal casts in \lean{e}. 
Based on the number and positions of applications of casts, we classify a lemma as
\begin{itemize}[itemindent=1.6em]
\itemsep1\jot
\item[\elim] if $\HC{lhs}=0$ and $\IC{lhs} \geq 1$
\item[\move] if $\HC{lhs} = 1$, $\IC{lhs} = \HC{rhs} = 0$, and $\IC{rhs} \geq 1$.
\item[\squash] if $\HC{lhs} \geq 1$, $\IC{lhs} = \IC{rhs} = 0$, and $\HC{lhs} > \HC{rhs}$.
\end{itemize}
An error is raised when 
a lemma does not fit in any of these categories is tagged with the \lean{norm_cast} attribute.
This classification applies to both \lstinline{=} and \lstinline{↔} lemmas. 
While users can override the classification by providing a parameter to the attribute, 
this is done for only 4 out of 424 attributions in \mathlib. 
Lean's user attribute API allows us to cache the set of classified lemmas
in a format convenient for the simplifier,
so the classifier creates very little overhead.

A previous version of \lean{norm_cast} relied on users classifying their lemmas manually.
However, misclassified lemmas can lead to errant behavior that is hard to diagnose.
From the perspective of library maintenance~\cite{Door20},
it is much cleaner to automatically classify the rewrite rules,
and we slightly redesigned the procedure to make this possible.

\section{Interface}
\label{section:api}

We provide a suite of tactics built around the core \lean{norm_cast} functionality. 
These try to replicate the behavior of other Lean tactics ``modulo casts,'' 
so that users can use familiar idioms while ignoring the presence of casts.

The core tactic \lean{norm_cast} simplifies the current goal. 
Alternatively, \lean{norm_cast at h} simplifies the type of a hypothesis \lean{h}. 
A variant \lean{exact_mod_cast t} 
simplifies both the goal and the type of the expression \lean{t}, 
and tries to use \lean{t} to close the goal; 
\lean{apply_mod_cast t} does similar, 
but allows arguments to \lean{t} to generate new subgoals. 
To close the goal with a hypothesis in the local context, 
\lean{assumption_mod_cast} will try \lean{exact_mod_cast} on all plausible candidates. 
Finally, \lean{rw_mod_cast [l₁, ..., lₙ]} will use a list of lemmas 
to sequentially rewrite the goal, 
calling \lean{norm_cast} in between rewrite steps. 
This generalizes the behavior of Lean's standard rewrite tactic \lean{rw}.

We also add \move and \squash lemmas into a custom simp lemma collection 
and define a tactic \lean{push_cast} that simplifies with this collection; 
note that \lean{push_cast} does not directly use the \lean{norm_cast} method. 
Calling \lean{push_cast} simplifies an expression in the opposite direction to \lean{norm_cast}, 
meaning that casts get pushed toward the leaf nodes of expressions. 
This does not allow casts to be eliminated over relations, but can be useful in its own right.

\section{Examples}
\label{section:examples}

The \lean{norm_cast} test file\footnote{\url{https://github.com/leanprover-community/mathlib/blob/master/test/norm_cast.lean}} 
in \mathlib demonstrates the tool in action. 
As a first example, we walk through a test where the heuristic splitting procedure is needed:
\begin{lstlisting}
n : ℕ, z : ℤ, h : ↑n - ↑z < (5 : ℚ)  ⊢  ↑n - z < (5 : ℤ)
\end{lstlisting}
Using \lean{exact_mod_cast h} will simplify \lean{h} to match the goal, 
which is already in normal form. 
After changing \lstinline{(5 : ℚ)} to \lstinline{↑(5 : ℕ)}, 
\lean{norm_cast} will fail to fire any \move or \elim rewrites. 
It will notice that \lstinline{↑n - ↑z} matches the heuristic splitting pattern, 
and rewrite \lstinline{↑n} to \lstinline{↑↑n}, 
where the inner cast goes \lstinline{ℕ → ℤ} and the outer goes \lstinline{ℤ → ℚ}. 
A \move rule will then match, rewriting the expression to \lstinline{↑(↑n - z) < ↑(5 : ℕ)}. 
While both sides of the \lean{<} are now casts to \lstinline{ℚ}, 
the left comes from \lstinline{ℤ} and the right from \lstinline{ℕ}, 
so no \elim rule will fire; instead, 
\lean{norm_cast} will match the entire expression to the heuristic splitting pattern 
and rewrite the right side to \lstinline{↑↑(5 : ℕ)}. 
It can then rewrite with an \elim lemma \lstinline{↑a < ↑b ↔ a < b} 
to obtain \lstinline{↑n - z < ↑(5 : ℕ)}, 
and finally normalize the numeral on the right to \lstinline{(5 : ℤ)}.

There is nothing special about the embedding domain $\QQ$ in the above example.
The theorem holds when \lean{n} and \lean{z} are embedded into any linear ordered ring.
\begin{lstlisting}
example (α : Type) [linear_ordered_ring α] (n : ℕ) (z : ℤ) (h : ↑n - ↑z < (5 : α)): 
  ↑n - z < (5 : ℤ) :=
by exact_mod_cast h
\end{lstlisting}

Lean's simplifier supports conditional rewriting,
and \lean{norm_cast} makes use of this support.
Note that the following example does not hold when \lean{n > m},
since \lean{m - n = 0}.
\begin{lstlisting}
example (m n : ℕ) (h : n ≤ m) : ↑(m - n) = (↑m - ↑n : ℤ) :=
by norm_cast
\end{lstlisting}  

The \lean{norm_cast} family of tactics is used throughout \mathlib. 
It is particularly useful in the development of the $p$-adic numbers $\Qp$ 
and integers $\Zp$~\cite{Lewi19}. 
The rationals $\QQ$ are embedded in the $p$-adics, 
and the definition of $\Qp$ requires working with a natural number $p$ embedded in $\ZZ$ and $\QQ$; 
furthermore, $\Zp$ is a subtype of $\Qp$. 
This development makes 64 calls to tactics in the \lean{norm_cast} family.

A lemma in the development of $\Qp$ bounds the $p$-adic norm of an integer:
\begin{lstlisting}
lemma le_of_dvd {n : ℕ} {z : ℤ} (hd : ↑(p^n) ∣ z) : 
  padic_norm p ↑z ≤ ↑p ^ -(↑n : ℤ)
\end{lstlisting}
The \mathlib proof of this lemma calls \lean{exact_mod_cast} four times,
to close subgoals:
\begin{itemize}
  \item \lstinline{0 ≤ p ⊢ 0 ≤ ↑p}
  \item \lstinline{1 ≤ p ⊢ 1 ≤ ↑p} 
  \item \lstinline{↑(p ^ n) ∣ z ⊢ ↑p ^ n ∣ z}
  \item \lstinline{↑z ≠ 0 ⊢ z ≠ 0}
\end{itemize}
The proof originally written without \lean{norm_cast} 
contains five explicit references to cast lemmas, 
and uses an explicit intermediate step that is unnecessary in the \mathlib proof:
\begin{lstlisting}
have hp' : (↑p : ℚ) ≥ 1, from 
  show ↑p ≥ ↑(1 : ℕ), from cast_le.2 (le_of_lt hp.gt_one)
\end{lstlisting}

The tool is particularly useful alongside the \lean{lift} tactic, 
which conditionally embeds terms in other types. 
In the following library lemma about the extended nonnegative reals \lean{ennreal}, 
lifting two \lean{ennreal}s to the type of nonnegative reals 
is justified by hypotheses that they are not infinite. 
In the resulting goal
\begin{lstlisting}
a b : nnreal ⊢ ennreal.to_real ↑a ≤ ennreal.to_real ↑b ↔ ↑a ≤ ↑b}
\end{lstlisting}
the casts on the left are \lean{nnreal → ennreal};
the goal is discharged immediately by \lean{norm_cast}.
\begin{lstlisting}
lemma to_real_le_to_real {a b : ennreal} (ha : a ≠ ⊤) (hb : b ≠ ⊤) : 
  ennreal.to_real a ≤ ennreal.to_real b ↔ a ≤ b :=
by { lift a to nnreal using ha, lift b to nnreal using hb, norm_cast }
\end{lstlisting}

Buzzard, Commelin, and Massot use \lean{norm_cast} 53 times 
in their definition of a perfectoid space~\cite{BCM20}. 
A typical use case is to match hypotheses from the \lean{units} subtype of a monoid 
to goals stated in the monoid itself, e.g.:
\begin{lstlisting}
γ γ₀ : units (Γ₀ R), h : γ₀ * γ₀ ≤ γ  ⊢  ↑γ₀ * ↑γ₀ ≤ ↑γ
\end{lstlisting}
These goals are rarely provable by conversion
because of the complicated algebraic structure on the types involved.
While traditional formalizations often make design decisions to limit the presence of coercions, 
they seem to be unavoidable in deep mathematical formalizations. 
Buzzard, Commelin, and Massot write that \lean{norm_cast} ``greatly alleviates \ldots pain'' in their project.

The performance of \lean{norm_cast} is not a major concern:
it depends mainly on the speed of Lean's simplifier,
which is heavily optimized 
and called regularly on complex goals.
Since \lean{norm_cast} calls the simplifier with a restricted set of rewrite rules,
it normally sees close to best-case performance.
It is very uncommon in practice 
to see \lean{norm_cast} take more than a fraction of a second to simplify an expression.
The rare cases in which it does are cases where other tactics,
including full and definitional simplification,
also tend to struggle.
We do not know of any benchmark suites for this type of problem 
and have not done extensive performance testing.

\section{Conclusion}
\label{section:conclusion}

The \lean{norm_cast} family of tactics can be seen as a variant of simplification procedures, 
which are common tools in proof assistants. 
Indeed, \lean{push_cast} is a straightforward application of Lean's simplifier, 
and similar functionality is found in many other systems, 
often in the default set of simplification lemmas.

Isabelle's standard simplifier~\cite{Nipk02} is more powerful than Lean's, 
but to our knowledge, 
the system has no tool like \lean{norm_cast}. 
Some theories may set up simp lemmas in a style that approximates our procedure, 
particularly for use with \lean{transfer}~\cite{Huff13}.
But it appears that approaches to managing and eliminating casts 
tend to be ad hoc combinations of simplification and manual work.

In Coq, unification hints~\cite{Asp09} can sometimes help 
to unify terms that differ in the placement of coercions.
When the necessary definitional equalities hold,
the behavior of \lean{assumption_mod_cast} and \lean{exact_mod_cast}
can be replicated by computation.
The same is true in Lean, 
and one benefit of \lean{norm_cast} is that it works even in the absence of definitional equalities,
as is often the case with abstract algebraic structures.
It is difficult to replicate the behavior of \lean{rw_mod_cast} using computation 
without providing more explicit information.
The \texttt{ppsimpl} preprocessing tool~\cite{Bess17}, 
which tries to eliminate inconvenient types and constants from a goal,
shares some design features with \lean{norm_cast}. 

The \lean{norm_cast} family aims to eliminate a source of frustration 
found when formalizing mathematical topics. 
The metaprogramming features of Lean allow it to be implemented in a lightweight and extensible way.
Its development was inspired by discussion between mathematical formalizers and tactic writers. 
We hypothesize that there are many other similarly lightweight tools 
that would help to move proof assistants closer to the mathematical vernacular.

The tool is inherently coupled to its ambient library,
meaning that it is only effective when the proper lemmas are tagged for its use. 
We thus consider it a mistake to consider tactic writing and library development separately. 
The \lean{norm_cast} tool and its corresponding lemma attributions are part of \mathlib, 
and despite not being themselves definitions or proofs, 
they constitute a different, procedural, kind of mathematical knowledge.

\paragraph{Acknowledgments.}
We thank Jasmin Blanchette for helpful comments,
Gabriel Ebner for contributions to the code,
and the \mathlib community for extensive stress-testing.
We thank an anonymous reviewer for enlightening examples
of how such problems can be dealt with in other systems.

%
%
%
\bibliographystyle{plainurl}
\bibliography{mathlib-paper}

\end{document}